\documentclass[12pt,preprint,onecolumn,longbibliography]{revtex4-2}
\usepackage{setspace}
\usepackage{graphicx}
\usepackage{amsmath} % The math mode package
\usepackage{mathtools}
\usepackage{amsfonts}
\usepackage{bbm}
\usepackage{braket}
\usepackage{caption}
\usepackage{graphicx}
\usepackage{float}
\usepackage{caption}
\usepackage{subcaption}
\usepackage{url}
\usepackage{color}
\usepackage[ruled]{algorithm2e}

\usepackage{amssymb}
\usepackage{cancel}
\usepackage{upgreek}
\usepackage{tabularx}
\usepackage{color}
\usepackage{units}
\usepackage{hhline}
\usepackage{textfit} %\scaletoheight{0.05cm}{very small text}
\usepackage{slashed}
\graphicspath{ {images/} }
\usepackage{graphicx}% Include figure files
\usepackage{dcolumn}% Align table columns on decimal point
\usepackage{bm}% bold math
\usepackage{hyperref}% add hypertext capabilities
\usepackage[mathlines]{lineno}% Enable numbering of text and display math
\usepackage{tikz}
\usetikzlibrary{positioning}
\usepackage{qcircuit}
\usepackage{braket}
\usepackage{indentfirst}
\usepackage{float}
\begin{document}

%\preprint{APS/123-QED}

\title{Geometrical picture of the electron-electron correlation at the large-D limit}
% Force line breaks with \\
%\thanks{A footnote to the article title}%
%\title{Dimensional Scaling for Quantum Criticality on a Quantum Computer}
%\maketitle
\vspace{-2em}

%%%%%%%%%%%%%%%%%%%%%%%%%%%%%%%%%
%\author{Kumar J. B. Ghosh\textsuperscript{1} Sabre Kais \textsuperscript{2} Dudley. R. Herschbach\textsuperscript{3} \endgraf
%\scriptsize\itshape \textsuperscript{1}Department of Electrical and Computer Engineering, University of Denver, Denver, CO, 80210, USA. \\ \textsuperscript{2} Chemistry department and Physics, Purdue University, West Lafayette, IN, 47906, USA.\\ \textsuperscript{3} Department of Chemistry and Chemical Biology, Harvard University  Cambridge MA 02138, USA.}
%
\author{Kumar J. B. Ghosh}
\email{jb.ghosh@outlook.com}
\affiliation{E.ON Digital Technology GmbH,\\ 
45131, Essen, Germany.
}%
\author{Sabre Kais}
\email{kais@purdue.edu}
\affiliation{ Department of Chemistry  and Physics,\\
Purdue University, West Lafayette, IN, 47906, USA.
}%
\author{Dudley R. Herschbach}
\email{dherschbach@gmail.com}
\affiliation{ Department of Chemistry and Chemical Biology,\\ Harvard University, Cambridge MA 02138, USA.
}%

%\nopagebreak

%\author{Kumar J B Ghosh, Teng Bian, Vivek Dixit, Sabre Kais}
%\email{kais@purdue.edu}
%\affiliation{%
%Chemistry department and Physics, Purdue University,\\
% West Lafayette, IN, 47906, USA.
%}%

%\collaboration{MUSO Collaboration}%\noaffiliation

%%%%%%%%%%%%%%%%%%%%%%%%%%%%%%%%%%%%%%%%%%%%%%%%%%%%%%%%%%%%%%%%%%%%%%%%%%%%%

%\author{Dudley. R. Herschbach}
%% %\homepage{http://www.Second.institution.edu/~Charlie.Author}
%\affiliation{
% Chemistry department, Harvard University
% }
%Authors' institution and/or address\\
 %This line break forced with \textbackslash\textbackslash
%}%

%\collaboration{CLEO Collaboration}%\noaffiliation

%\date{}% It is always \today, today,
             %  but any date may be explicitly specified
%\maketitle

\begin{abstract}
In electronic structure calculations, the correlation energy is defined as the difference between the mean field and the exact solution of the non relativistic Schr\"odinger equation. Such an error in the different calculations is not directly observable as there is no simple quantum mechanical operator, apart from correlation functions,  that correspond to such quantity. Here, we use the dimensional scaling approach, in which the electrons are localized at the large-dimensional scaled space, to describe a geometric picture of the electronic correlation.  Both, the mean field, and the exact solutions at the large-D limit have distinct geometries. Thus, the difference might be used to describe the correlation effect. Moreover, correlations can be also described and quantified  by the entanglement between the electrons, which is a strong correlation without a classical analog. Entanglement is directly observable and it is one of the most striking properties of quantum mechanics and bounded by the area law for local gapped Hamiltonians of interacting many-body systems. This study opens up the possibility of presenting a geometrical picture of the electron-electron correlations and might give a bound on the correlation energy. The results at the large-D limit and at D$=3$ indicate the feasibility of using the geometrical picture to get a bound on the electron-electron correlations.
\end{abstract}
%\pacs{Valid PACS appear here}% PACS, the Physics and Astronomy
                             % Classification Scheme.
%\keywords{large-D limit,  metallic hydrogen, correlation energy, entanglement entropy, area law.}%Use showkeys class option if keyword                           %display desired
\maketitle

%%%%%%%%%%%%%%%%%%%%%%%%%%%%%
%\tableofcontents
\newpage
%%%%%%%%%%%%%%%%%%%%%%%%%%%

%\tableofcontents

%\doublespacing

\section{\label{sec: Introduction}Introduction}

Dimensional scaling, as applied to electronic structure calculations, offers promising computational strategies and heuristic perspectives with physical insight of the electronic structure of atoms, molecules and extended systems \cite{Stillinger1975,MLODINOW1980314,yaffe1982large,Herschbach1986}. Taking the  spatial dimension of the physical space as a variable other than $D=3$ can make a problem much simpler and then one can  use perturbation theory or other techniques to obtain the  results at $D=3$. The  $D$-scaling technique was used first in quantum chromodynamics \cite{witten1980} and then applied to the  Helium atom\cite{MLODINOW1980314,yaffe1982large,Herschbach1986}. In this approach, we solve the problem at the $D \to \infty$ limit and then add terms in powers of $\delta=1/D$. Using different  summation techniques \cite{goodson1992large} can obtain highly accurate results for $D=3$. The  dimensional scaling approaches were  extended to $N$-electron atoms \cite{Zhen1993}, renormalization with $1/Z$ expansions \cite{kais19941}, random walks \cite{rudnick1987shapes, doi:10.1021/acs.jpca.1c05551}, interpolation of hard sphere virial coefficients \cite{kais-hard-sphere}, resonance states \cite{kais-resonances} and dynamics of many-body systems in external fields \cite{kais-laser1,kais-laser2}. We refer the reader to the book 
 ``Dimensional scaling in chemical physics" \cite{herschbach2012dimensional} for more details of the approach and applications. 

In computational physics and chemistry, the Hartree–Fock (HF) method \cite{slater1951simplification} is a self-consistent field approximation to determine the wave function and the energy of a quantum many-body system in a stationary state. This method is based on the idea that we can approximately describe an interacting electronic system in terms of an effective single-particle model. Moreover, this simple approximation remains the starting point for more accurate post Hartree-Fock methods such  as  coupled clusters and configuration interactions. 

The Hartree–Fock method  assumes that the exact $N$-body wave function of the system can be approximated by a single Slater determinant of $N$ spin-orbitals. In quantum chemistry calculations, the correlation energy is defined as the difference between the Hartree–Fock
limit energy and the exact solution of the nonrelativistic Schr{\"o}dinger equation \cite{lowdin1958correlation}. Other measures of electron correlation also exist in the literature, for e.g. the statistical correlation coefficients \cite{kutzelnigg1968correlation}. Recently the Shannon entropy is also described as a measure of the correlation strength \cite{guevara2003shannon, shi2004finite}. Electron correlations have wide implications on atomic, molecular \cite{wilson2014electron}, and solid state physics \cite{march1999electron}. Observing the correlation energy for large systems is one of the most challenging
problems in quantum chemistry because there is no simple operator in quantum mechanics that its measurement gives the correlation energy. This leads to proposing the entanglement as an alternative  measure of the electron correlation for atoms and molecules\cite{huang2005entanglement}. All the information needed for quantifying the entanglement is contained in the two-electron density matrix. This measure is readily calculated by evaluating the von Neumann entropy of the one electron reduced density operator. As an example, one can see the  calculation of the entanglement for He atom and H$_2$ molecule with different basis sets\cite{huang2005entanglement}. The advantage of this proposal is that entanglement is directly observable, and it is one of the most striking properties of quantum mechanics.

Entanglement is a quantum mechanical property that describes  strong correlations between quantum mechanical particles that has no classical analog and has been studied extensively in the field of quantum information and quantum computing \cite{bennett2000quantum, macchiavello2000quantum,Nielsen,gruska1999quantum,schrodinger1935gegenwartige, bruss2002characterizing}. Moreover, scientists studied numerous properties of the entanglement entropy \cite{audenaert2002entanglement, osborne2002entanglement,osterloh2002scaling, vidal2003entanglement}, addressing many interesting topics of physics, for example black hole physics \cite{bekenstein2020black,bombelli1986quantum, srednicki1993entropy}, distribution of quantum correlations in quantum many-body systems in one dimension \cite{amico2008entanglement, audenaert2002entanglement, barthel2006entanglement, barthel2006entanglement} and higher dimensions \cite{bravyi2006lieb, cramer2006correlations, cramer2006entanglement, plenio2005entropy}, complexity of quantum many-body systems and
their simulation \cite{schollwock2005density, white1992density}, and topological entanglement entropy \cite{nussinov2009symmetry, wen1989vacuum, witten1989quantum, haque2007entanglement, kitaev2006topological}.

In this article we describe a geometric interpretation of correlation energy calculated at large-$D$ limit and at three dimensions and establish a relation between the correlation energy and the area law of entanglement.  In section \ref{sec:Area law of entanglement}, we describe the area law of entanglement.  In sections \ref{sec:correlation energy in the large-D limit} and \ref{sec:Electron correlation and the surface area for metallic hydrogen}, we describe the relation between the area difference and the correlation energy of the  atomic/ionic systems and metallic hydrogen at the large-$D$ limit.  In section \ref{sec:correlation energy in three dimension}, we consider the helium atom and the metallic hydrogen at $D=3$, where the electrons are not localized unlike in the $D \to \infty$ limit.  Finally, in section \ref{sec:Conclusion}, we make some concluding remarks. We adopt Hartree atomic units for our calculations.

\section{\label{sec:Area law of entanglement} Area law of entanglement}
In classical physics, concepts of entropy quantify the amount of information that is lacking to identify the microstate of a physical system from all the possibilities compatible with the macrostate of the system. In quantum mechanics, we define the entanglement entropy or geometric entropy \cite{eisert2010colloquium, horodecki2009quantum, plenio2014introduction}, which arises because of a very fundamental property called entanglement.  In quantum many-body systems, for a pure state $\rho = |\psi  \rangle  \langle \psi |$, the von Neumann entropy is a good measure of entanglement and defined as \cite{eisert2010colloquium} 
\begin{equation}
S(\rho) = - \text{tr} \left[ \rho \log_2 \rho \right]. 
\end{equation}

\begin{figure}[h]
    \centering
    \begin{tikzpicture}[
 image/.style = {text width=0.8\textwidth, 
                 inner sep=0pt, outer sep=0pt},
node distance = 1mm and 1mm
                        ] 
\node [image] (frame1)
    {\includegraphics[width=\linewidth]{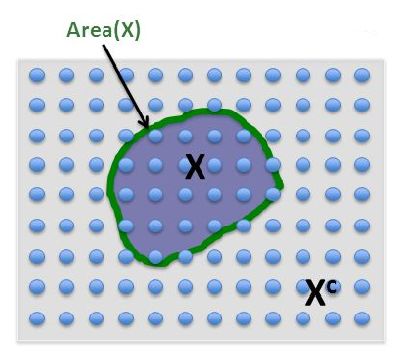}};
\end{tikzpicture}
   \caption{The area law of the entanglement entropy: We have n-qubits (spin $\frac{1}{2}$ particles) in two-dimensional space interacting with local Hamiltonians (sum over nearest neighbor integrating particle)  and the question is what is the entanglement between the interior shaded area $X$ and the exterior system $X^c$. The conjecture of the area law is that the entanglement is bounded by the size of the boundary as shown above. The number of qubits in Fig. \ref{area_law} is for the demonstration purpose only (just a cartoon).} 
    \label{area_law}
\end{figure}	

Generally,  a quantum state $|\psi  \rangle $	of	$n$-qubits (spin $\frac{1}{2}$ ) 	is	represented as a vector in $\left(\mathbb{C}^2\right)^{\bigotimes n}$, 

\begin{equation}
|\psi  \rangle = \sum_{i_1, ... , i_n}  C_{i_1, ... , i_n} |i_1, ... , i_n \rangle.
\end{equation}

This is a very complex wave function  with complex coefficients $C_i$, in Hilbert space of dimension   $\mathbb{C}^{2^{n}}$.  
One way to find a possible efficient representation, is to examine a bipartite system with a local gapped Hamiltonian, summing over nearest neighbor interacting particles, as shown in Fig. \ref{area_law}\cite{hastings2007area, hastings2007entropy, arad2013area, brandao2013area, landau5143polynomial}. The  entanglement  entropy between the interior state $X$ and the exterior state $X^c$  scales as the size of the boundary 	for	every	region $X$,
\begin{equation}
S(X) \leq \text{ constant} \times  \text{ Area}(X). \label{area_law_for_entanglement_entropy}		
\end{equation}	

Recently, the area law of bounding the entanglement entropy of the  ground state energy was examined for local integrating particles with large gapped energy spectrum  \cite{hastings2007area, hastings2007entropy, arad2013area, brandao2013area, landau5143polynomial}. It is shown that the ground state of a chain of $d$-dimensional spins with a boundary $L$ and spectral gap $\delta$ is  bounded by  an area law
\begin{equation}
S_{1D} (L) = \exp \left(\mathcal{O} \left(\log(d)/\delta\right)\right) = \text{ constant},
\end{equation} 
where $\mathcal{O}(...)$ is used to denote a bound up to a numeric constant of order unity.  
The area law for the entanglement entropy and ground state of local Hamiltonian was further extended to two dimensional lattice \cite{ wolf2008area, masanes2009area, de2010ground, michalakis2012stability, anshu2021area}. Here, we use the area law as a heuristic approach to guide us to discuss having a possible bound on  the electronic correlation  energy as one can obtain a well define geometrical picture of the localized electrons at the large-$D$ limit, for both the mean field and the exact electronic structure.

\section{\label{sec:correlation energy in the large-D limit} Correlation energy and the surface area for atomic systems in the large-D limit}

At the large-D limit, one can obtain a simple formula for the electronic structure of $N$-electron atoms \cite{Zhen1993}. At $D \to \infty$ limit,  the electrons are localized in the scaled space with equidistant $r_m$ from the nucleus and equiangular with respect to each other, which means that there is no shell structure. This simple geometrical picture  holds for a wide variety of atoms and ions including all neutral atoms with $Z < 14$ and all positive ions with $N/Z \lesssim 0.936$ \cite{Zhen1993}. In the Hartree-Fock limit all electrons are orthogonal to each other. The electronic energy $\epsilon_\infty^{HF}$ and distance from the nucleus $r_m^{HF}$ of an $N$-electron atom are described by the following formulas:
\begin{equation}
\epsilon_\infty^{HF} = -\frac{N}{2} \left[ 1- 2^{3/2} \left( N-1 \right) \lambda \right]^2, \label{energy_HF}
\end{equation}
\begin{equation}
r_m^{HF} = \left[ 1- 2^{3/2} \left( N-1 \right) \lambda \right]^{-1}, \label{rm_HF}
\end{equation} 
with $\lambda = 1/Z$, where $Z$ is the nuclear charge.  For larger atoms the global minimum of the Hamiltonian $H_\infty$ no longer has a maximal symmetry and the solutions display a kind of meta-shell structure, but one which is apparently unrelated to the normal three dimensional shell structure \cite{Zhen1993}.

After adding the inter-electronic correlation in the picture the electrons are no more orthogonal to one another. Therefore, the inter-electronic angle $\theta_\infty$ becomes slightly larger than $\pi/2$, although their equi-distance property from the nucleus $\rho_\infty$ still holds. At $D \to \infty$ limit the relevant quantities can be calculated by the following formula:

The exact  energy is given by
\begin{equation}
\epsilon_\infty = -\frac{1}{2}  \left( \frac{1- \xi}{1- \xi/N} \right)^3 \left( N - N \xi + \xi\right) , \label{energy_total}
\end{equation}
the inter-electronic angle
\begin{equation}
\theta_\infty = \arccos \left( \frac{\xi}{\xi - N} \right) , \label{theta_total}
\end{equation}
and the electronic distance
\begin{equation}
\rho_\infty  =  \left( \frac{1- \xi/N}{1- \xi} \right)^2 \left( N - N \xi + \xi\right) , \label{rm_total}
\end{equation}
where the parameter $\xi$ is the smallest positive root obtained by solving the following quartic equation
\begin{equation}
8 N Z^2 \xi^2 \left( 2- \xi \right) ^2 = \left( N- \xi \right)^3.
\end{equation}

We solve the above sets of equations (\ref{energy_HF}) and (\ref{energy_total}) numerically and calculate the correlation energy at $D \to \infty$ limit which is defined by 
\begin{equation}
\epsilon^{Corr}_\infty = \mid \epsilon_\infty - \epsilon_\infty^{HF} \mid. 
\end{equation}

For example, from the above equations ( \ref{energy_HF}, \ref{rm_HF}, \ref{energy_total},  \ref{theta_total}, \ref{rm_total}),  we obtain the following results for the helium-atom

$r_m^{HF} = 1.214737$ , $\epsilon_\infty^{HF} = -0.6776966$, $\rho_\infty = 1.213927$, 
$\theta_\infty = 1.663309$ rad, $\epsilon_\infty = -0.68444228$, and $\epsilon^{Corr}_\infty = 0.0067456$. 

\begin{figure}[h]
    \centering
    \begin{tikzpicture}[
 image/.style = {text width=0.49\textwidth, 
                 inner sep=0pt, outer sep=0pt},
node distance = 1mm and 1mm
                        ] 
\node [image] (frame1)
    {\includegraphics[width=\linewidth]{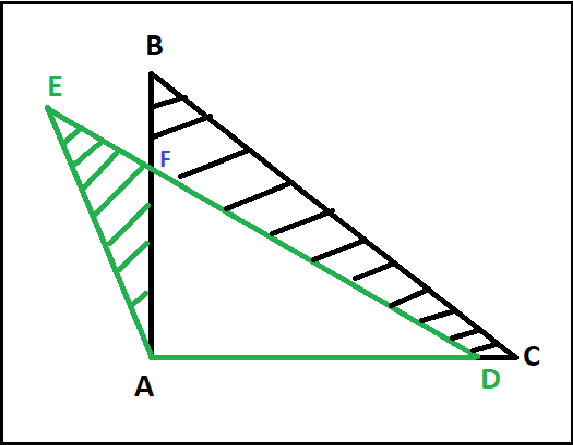}};
\node [image,right=of frame1] (frame2) 
    {\includegraphics[width=0.89\linewidth]{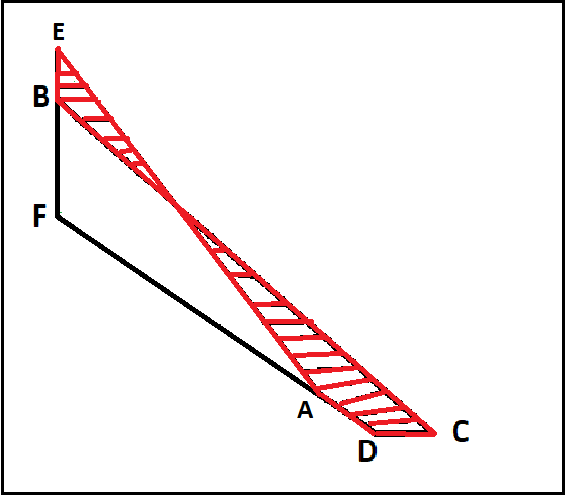}};
\end{tikzpicture}
    \caption{The geometry of the localized electrons in a helium atom is described at large-$D$ limit. Without inter-electronic correlation the electrons $B$ and $C$ are orthogonal with respect to the nucleus $A$. Whereas, with correlation the inter-electronic angle $\angle DAE$ becomes slightly larger than $\pi/2$. At the left panel, we draw the triangles formed by the two electrons and the nucleus with ($\bigtriangleup CAB$) and without correlation ($\bigtriangleup DAE$). At the right, the area difference between the two triangles is described by the striped red region.} 
    \label{figure_triangle}
\end{figure}

Next, we draw  the following geometrical pictures of the two localized electrons of the  helium-atom at $D \to \infty$. We first construct a right angle triangle with two equal sides equal to $r_m^{HF}$ and another isosceles triangle with two equal sides equal to $\rho_\infty$ and the inter-electronic angle $\theta_\infty$.

In Fig. \ref{figure_triangle}, the nucleus is situated at the point $A$. The points $B$ and $C$ describe the positions of the two electrons at HF-limit whereas the points $D$ and $E$ describe the positions of the two electrons with correlation. We see that at HF-limit $AB \perp AC$ and with correlation $\angle EAD > \pi/2$. We calculate the areas of an  isosceles triangle with the formula $\bigtriangleup ADE = \frac{1}{2} AE \times AD \times \sin (\angle EAD)$. Then we calculate the magnitude of the area difference ($\bigtriangleup \text{area}$) between the $\bigtriangleup ABC $ and $\bigtriangleup ADE$  for the helium atom in the following:
$\bigtriangleup \text{ area}  = 0.00413417$. In the figure the area difference is the difference between the stripped areas.

\begin{figure}[h]
    \centering
    \begin{tikzpicture}[
 image/.style = {text width=0.5\textwidth, 
                 inner sep=0pt, outer sep=0pt},
node distance = 1mm and 1mm
                        ] 
\node [image] (frame1)
    {\includegraphics[width=\linewidth]{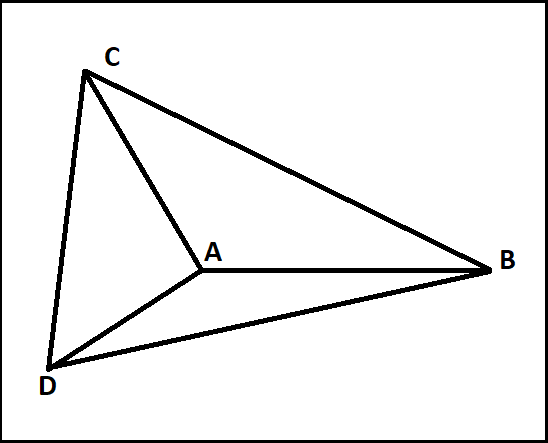}};
\end{tikzpicture}
   \caption{The above figure describes the geometry of a three-electron atom at $D \to \infty$ limit.} 
    \label{figure_triangle_2}
\end{figure} 

For three-electron atoms we construct the three triangles $\bigtriangleup ABC$, $\bigtriangleup ACD$, and $\bigtriangleup ADB$ (see Fig. \ref{figure_triangle_2}). The point $A$ is the position of nucleus and the points $B, C, D$ are three localized electronic  positions. The sides $AD= AC = AB$. At HF-limit, the angles  $\angle BAC = \angle CAD = \angle BAD = \pi/2$, whereas, with correlation $\angle BAC = \angle CAD = \angle BAD > \pi/2$.

As an extension of the above, we construct $N$-triangles for $N$-electron atoms and compute the total areas for both HF and with inter-electronic correlation. Then we compute the area difference for $N$-triangles between HF and electronic-correlation.

\begin{figure}[h]
    \centering
    \begin{tikzpicture}[
 image/.style = {text width= 0.95\textwidth, 
                 inner sep=0pt, outer sep=0pt},
node distance = 1mm and 1mm
                        ] 
\node [image] (frame1)
    {\includegraphics[width=\linewidth]{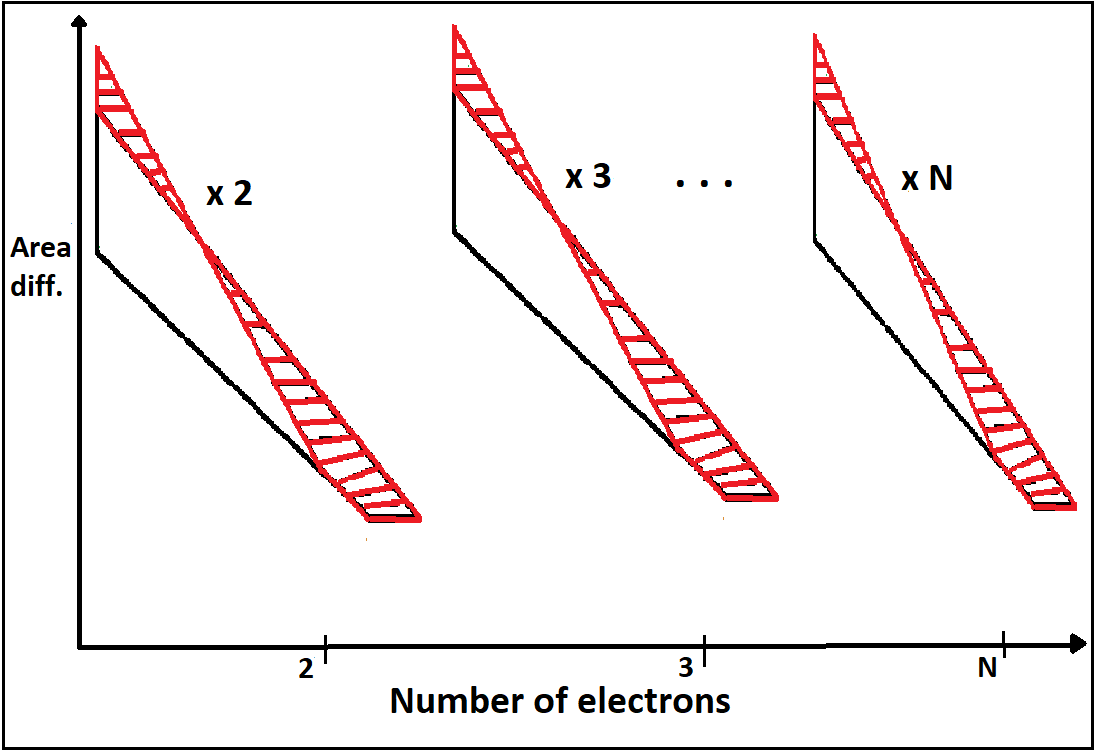}};
\end{tikzpicture}
   \caption{The above figure describes the area difference calculated for a $N$-electron atom at $D \to \infty$ limit.} 
    \label{figure_triangle_3}
\end{figure} 

%We compute all the relevant physical quantities and parameters for neutral atoms for $N=2$ to $N=14$ by the formulas described above. 
In Fig. \ref{figure_triangle_3}, we describe the area difference which is computed from the $N$-triangles of the $N$-electron atom.

\begin{figure}[h]
    \centering
    \begin{tikzpicture}[
 image/.style = {text width=0.8\textwidth, 
                 inner sep=0pt, outer sep=0pt},
node distance = 1mm and 1mm
                        ] 
\node [image] (frame1)
    {\includegraphics[width=\linewidth]{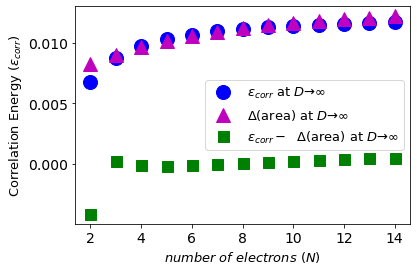}};
\end{tikzpicture}
   \caption{In the above figure we plot the electronic correlation energies for neutral atoms from $N=2$ to $N=14$ in blue, and the area difference described above in purple. The green points are the difference between the above two results.} 
    \label{figure_correlation}
\end{figure}

We compute the correlation energies for $N$-electron atoms, using equations (\ref{energy_HF}, \ref{energy_total}), and plot this in Fig. \ref{figure_correlation} along with the area differences obtained above.

\begin{figure}[h]
    \centering
    \begin{tikzpicture}[
 image/.style = {text width=0.8\textwidth, 
                 inner sep=0pt, outer sep=0pt},
node distance = 1mm and 1mm
                        ] 
\node [image] (frame1)
    {\includegraphics[width=\linewidth]{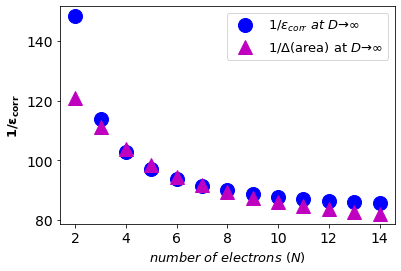}};
\end{tikzpicture}
   \caption{In the above figure we plot the inverse of the electronic correlation energies for neutral atoms from $N=2$ to $N=14$ in blue, and the inverse of area difference in purple. } 
    \label{figure_inverse_correlation}
\end{figure}

We also compute the inverse of the correlation energies and  the inverse of the area difference obtained from the above prescription and plot it in Fig. \ref{figure_inverse_correlation}.
 
In figures (\ref{figure_correlation}, \ref{figure_inverse_correlation}), we see that the area difference  is a  close estimate to the correlation energy at the large-$D$ limit. On the other hand, it was shown by Loeser et al. \cite{Zhen1993} that the correlation energy at $D=3$ is a good approximation to the correlation energy at $D \to \infty$. Therefore the correlation energy is bounded by the area difference of the electronic triangles between the HF-limit and with correlation at large-$D$ limit.  We plot the known accurate correlation energies at $D=3$ \cite{davidson1991ground, veillard1968correlation, Herschbach2017, kais1994large} and with the correlation energies obtained at $D=\infty$.

\begin{figure}[h]
    \centering
    \begin{tikzpicture}[
 image/.style = {text width=0.8\textwidth, 
                 inner sep=0pt, outer sep=0pt},
node distance = 1mm and 1mm
                        ] 
\node [image] (frame1)
    {\includegraphics[width=\linewidth]{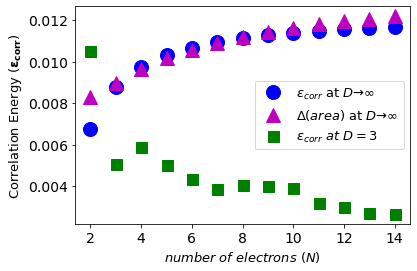}};
\end{tikzpicture}
   \caption{In the above figure we plot the electronic correlation energies for neutral atoms from $N=2$ to $N=14$ in blue, the area difference in purple, and correlation energy at $D=3$ result in green. Note that the energies are represented in the $Z^2$ hartree units. } 
    \label{figure_D3}
\end{figure}

From figure \ref{figure_D3}, we see that the correlation energies at $D=3$ is bounded by the area difference at $D \to \infty$ limit, only exception for $N=2$. 

%Note that the energies are represented in the $Z^2$ hartree units.
%
%For most of the calculation at large-$D$ limit we define the energy in $Z^2$ hartree unit. Although, in articles \cite{davidson1991ground, veillard1968correlation} the correlation energy  is described in hartree unit. So we compare the $Z^2 \times$ (area difference) and the correlation energy in atomic unit in the following figure.
%
%\begin{figure}[H]
%    \centering
%    \begin{tikzpicture}[
% image/.style = {text width=0.8\textwidth, 
%                 inner sep=0pt, outer sep=0pt},
%node distance = 1mm and 1mm
%                        ] 
%\node [image] (frame1)
%    {\includegraphics[width=\linewidth]{pictures/3dresult_z2.png}};
%\end{tikzpicture}
%   \caption{In the above figure we plot the electronic correlation energies for neutral atoms from $N=2$ to $N=14$ in blue, the $Z^2 \times$ (area difference) in purple, and correlation energy at $D=3$ result in green. Note that the energies are represented in the hartree units. } 
%    \label{figure_D3z2}
%\end{figure}
%
%From the above figure \ref{figure_D3z2}, we  see that the correlation energies at $D=3$ is bounded by the area difference at $D \to \infty$ limit.

\section{\label{sec:Electron correlation and the surface area for metallic hydrogen}Electron correlation and the surface area for metallic hydrogen at the large-D limit}
In 1935, Wigner and Huntington predicted the metallization of hydrogen\cite{WignerandHuntington}, a phase of hydrogen that behaves like an electrical conductor. Ever since this has been a major quest for condensed matter physics, pursuing theory \cite{loeser1993large, mao1994ultrahigh, zha2012synchrotron, dias2017observation, mcminis2015molecular} and extreme high-pressure experiments \cite{hemley1988phase, hemley1991high, hemley1996synchrotron, zaghoo2016evidence, Silvera_2018, loubeyre2020synchrotron}. Dimensional scaling and interpolation as applied to metallic hydrogen was investigated in articles \cite{loeser1993large, metallic_hydrogen}. With appropriate scaling, energies will be in units of $4/(D-1)^2$ hartrees, and distances in units of $D(D-1)/6$ Bohr radii. we consider the lattice to be simple cubic (SC).  With scaling and simplifications, the Hartree-Fock one-electron Hamiltonian in the $D\to\infty$ limit in a lattice of hydrogen atoms with clamped nuclei can be written as \cite{loeser1993large}:

%With the appropriate scaling, described in the previous section (\ref{sec:Dimensional scaling formula for extended systems}),  the Hartree-Fock one-electron Hamiltonian in the $D \to \infty$ limit in a lattice of hydrogen atoms with clamped nuclei can be written as \cite{loeser1993large}:
\begin{equation}
\mathcal{H} = \frac{9}{8 \rho^2} -  \frac{3}{2 \rho} + W(\rho,R), \label{Hamiltonianmetallichydrogen}
\end{equation}
where
\begin{equation}
W(\rho,R) = \frac{3}{4} \sum_{l,m,n \in \mathcal{L}^\prime}\frac{1}{\sqrt{\sigma^2R^2}} -\frac{2}{\sqrt{\sigma^2R^2 + \rho^2}}+\frac{1}{\sqrt{\sigma^2R^2 + 2\rho^2}},
\end{equation}
with
\begin{equation}
\sigma^2 = l^2+m^2+n^2 .
\end{equation}

For any specified lattice type and scaled lattice constant $R$, the minimum of Eq. (\ref{Hamiltonianmetallichydrogen}) with respect to $\rho$ gives the energy per electron.  The whole lattice is three-dimensional, noted $\mathcal{L}^\prime$ minus the one site $(0,0,0)$.  The single variable $\rho$ is the orbit radius and $R$ is the lattice spacing.  

We introduce the inter-electronic correlation at $D \to \infty$ limit by opening up the dihedral angles from their Hartree-Fock values of exactly $\pi/2$ rad. The dihedral angles in the correlated solution is determined by two effects, namely, the centrifugal effects, favoring $\pi/2$ rad, and interelectron repulsions, favoring $\pi$ rad. Although the final effect turns out to be the angles very close to $\pi/2$ rad. For the calculation purpose we assume the inter-electronic correlation is upto third nearest neighbor, which is a very legitimate assumption. We assume the lattice structure to be simple cubic (SC).

At $D\to \infty$ limit, the Hamiltonian with inter-electronic correlation can be written as \cite{loeser1993large}:
\begin{equation}
\mathcal{H}_{corr} = \mathcal{H}_{HF}+ \frac{9}{8 \rho^2}\left( \left(\frac{\Gamma^\prime}{\Gamma}\right)^{(3)} -1 \right) +  \frac{3}{2}  W_\Delta^{(3)}, \label{Hamiltonianmetallichydrogenwithcorrelation}
\end{equation}
where
\begin{equation}
W_\Delta^{(3)} = 6 \Delta W(\rho, R, \gamma_{100})+ 12 \Delta W(\rho, \sqrt{2} R, \gamma_{110})+ 8 \Delta W(\rho, \sqrt{3} R, \gamma_{111}), 
\end{equation}
with
\begin{equation}
W(\rho, \sigma R, \gamma_{lmn}) = \frac{1}{2} \left(\frac{1}{\sqrt{\sigma^2R^2 + 2 \rho^2 \left( 1- \gamma_{lmn} \right)}} - \frac{1}{\sqrt{\sigma^2R^2 + 2\rho^2}}\right).
\end{equation}
%and
%\begin{equation}
%\sigma^2 = l^2+m^2+n^2 .
%\end{equation}

\begin{figure}[H]
    \centering
    \begin{tikzpicture}[
 image/.style = {text width=0.5\textwidth, 
                 inner sep=0pt, outer sep=0pt},
node distance = 1mm and 1mm
                        ] 
\node [image] (frame1)
    {\includegraphics[width=\linewidth]{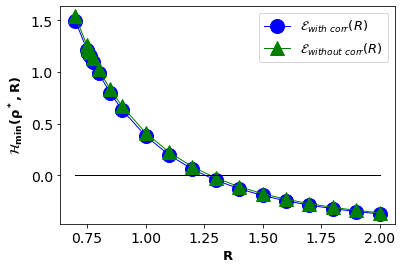}};
\node [image,right=of frame1] (frame2) 
    {\includegraphics[width=\linewidth]{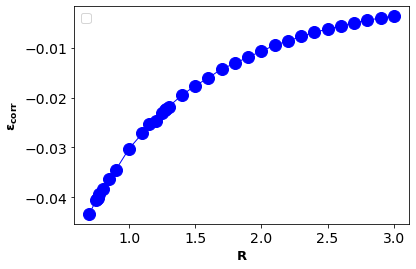}};
\end{tikzpicture}
\caption{In the left, we plot the energy obtained at $D \to \infty$ with inter-electronic correlation in blue, compared with the HF energy as a function of $R$ in green. In the right, we plot the correlation energy as a function of $R$.  }
\label{correlation_energy_inf}
\end{figure}

The $\mathcal{H}_{HF}$ is the Hamiltonian in Hartree-Fock approximation defined in Eq. (\ref{Hamiltonianmetallichydrogen}). 
The quantity  $\gamma_{lmn} = \cos \theta_{lmn}$, with $\theta_{lmn}$ are the dihedral angles which is very close to $\pi/2$.  The Gramian ratio $\left( \frac{\Gamma^\prime}{\Gamma}\right)^{(3)}$ is defined as \cite{loeser1993large}
\begin{equation}
\left( \frac{\Gamma^\prime}{\Gamma} \right)^{(3)} \simeq 1 + 6 \gamma_{100}^2 + 12 \gamma_{110}^2 + 8 \gamma_{111}^2 + (\text{higher order terms in } \gamma).
\end{equation} 

We optimize the above Hamiltonian (\ref{Hamiltonianmetallichydrogenwithcorrelation}) with respect to the parameters $\gamma_{100}, \gamma_{110}, \gamma_{111}$ keeping the values of $\rho$ and $R$ from the HF-Hamiltonian \cite{metallic_hydrogen}. In Fig. \ref{correlation_energy_inf},  we plot the minimum values of the Hamiltonian (\ref{Hamiltonianmetallichydrogenwithcorrelation}) at $D \to \infty$ limit as function of $R$ and  compare with the values obtained from the HF-Hamiltonian. We also plot the correlation energies $\mathcal{E}_{corr}=\mathcal{H}_{corr}-\mathcal{H}_{HF}$  as a function of $R$.

%$\rho^\star = 0.932$ and 
In the left hand side of Fig. \ref{correlation_energy_inf}, we see that the ground state energy becomes positive for $R < 1.28$, therefore makes the system is unstable. Therefore, $R > 1.28$ can be think of a physically stable region for MH at $D \to \infty$ limit. 
%Therefore, at the point  $R = 1.28$, the system goes through a phase transition. 

In simple cubic lattice at $D \to \infty$ limit the electrons also forms a cubic structure  as follows. With each reference electron there are $6$ nearest neighbors at a distance $R$, $12$ second nearest neighbors at a distance $\sqrt{2} R$, $8$ third nearest neighbors  at a distance $\sqrt{3} R$ and so on. Now, at HF-limit the nearest neighbours ($N_{1i}$ for $i =1,2, ... 6$) are orthogonal to each other with respect to the reference electron ($O$). For correlation the dihedral angles between the electrons becomes slightly greater than $90^{\circ}$. The following figure describes a cross section of a MH in SC lattice with reference electron $O$. In HF limit we consider $6$ square surfaces for $6$ neighbouring electrons at a distance $R$, $12$ square surfaces for $12$ neighbouring electrons at a distance $\sqrt{2} R$, and $8$ square surfaces for $8$ neighbouring electrons at a distance $\sqrt{3} R$. Whereas, for with inter-electronic correlation the each square becomes a rhombus. 

\begin{figure}[h]
    \centering
    \begin{tikzpicture}[
 image/.style = {text width=0.8\textwidth, 
                 inner sep=0pt, outer sep=0pt},
node distance = 1mm and 1mm
                        ] 
\node [image] (frame1)
    {\includegraphics[width=\linewidth]{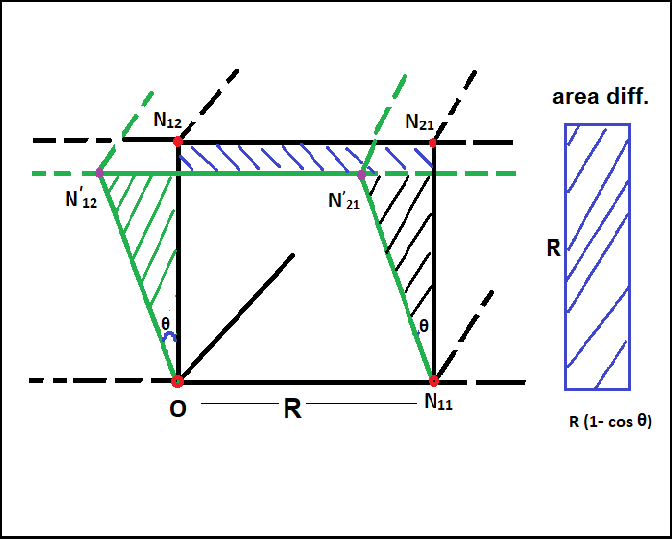}};
\end{tikzpicture}
   \caption{The area difference for a single cell formed by each nearest neighbor.} 
    \label{area_diff_MH}
\end{figure}

In figure \ref{area_diff_MH}, we plot the area difference due to the electronic correlation. The area difference between the square and the rhombus formed by the each nearest neighbour is equal to $R^2 (1-\cos \theta)$, where $\theta = \gamma_{100}-90^{\circ}$ is the angle deviation from $90^{\circ}$ due to correlation effect. The area difference for each next nearest neighbor is equal to $2 R^2 (1-\cos \theta_2)$, with $\theta_2 = \gamma_{110}-90^{\circ}$, and so on. The total area difference due to the correlation upto the third nearest neighbor is given by:
\begin{equation}
\Delta \text{area} = 6 R^2 \left( 1- \cos (\gamma_{100}-90^{\circ}) \right) + 24 R^2 \left( 1- \cos (\gamma_{110}-90^{\circ}) \right) + 24 R^2 \left( 1- \cos (\gamma_{111}-90^{\circ}) \right).
\end{equation}

We vary the lattice parameter $R$ and calculate the area difference for each value of $R$.  In the following figure\ref{figure_corr_ene_and_area_diff_MH}, we plot the correlation energy per electron for MH in SC lattice at large-$D$ limit and the area difference described above as a function of the lattice parameter $R$.

\begin{figure}[h]
    \centering
    \begin{tikzpicture}[
 image/.style = {text width=0.8\textwidth, 
                 inner sep=0pt, outer sep=0pt},
node distance = 1mm and 1mm
                        ] 
\node [image] (frame1)
    {\includegraphics[width=\linewidth]{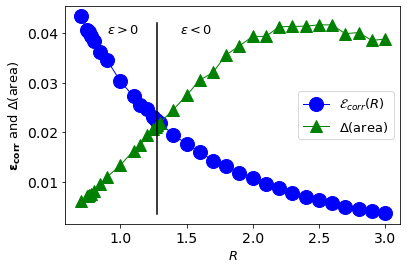}};
\end{tikzpicture}
   \caption{We plot the correlation energy per electron in MH at $D \to \infty$ as a function of $R$ in blue and the area difference in green. } 
    \label{figure_corr_ene_and_area_diff_MH}
\end{figure}

In Fig. \ref{figure_corr_ene_and_area_diff_MH}, we see that the correlation energy per electron is bounded by the area difference in the physically stable region (i.e. total ground state energy $\epsilon > 0$).

The correlation energy per electron (in Rydberg unit) in metallic hydrogen at $D=3$ was calculated by Neece et al. \cite{NEECE1971621, ross1977molecular}
\begin{equation}
\epsilon_{corr} = - 0.1303 + 0.0495 ~\ln (r_s) . \label{Neece}
\end{equation}

In the above Eq. (\ref{Neece}), the lattice constant $R$ is related to $r_s$, the standard solid state parameter, defined as the radius of a sphere (in $a_0$  bohr units) in which contains on average one electron. For the SC lattice, 
\begin{equation}
\frac{4}{3} \pi r_s^3 = R^3 . \label{rs_simple_cubic}
\end{equation}

\begin{figure}[H]
    \centering
    \begin{tikzpicture}[
 image/.style = {text width=0.8\textwidth, 
                 inner sep=0pt, outer sep=0pt},
node distance = 1mm and 1mm
                        ] 
\node [image] (frame1)
    {\includegraphics[width=\linewidth]{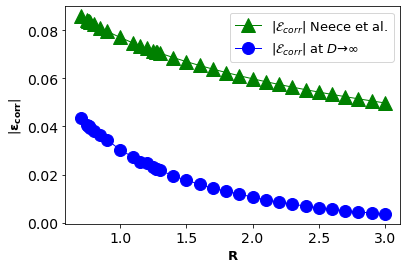}};
\end{tikzpicture}
   \caption{We plot the correlation energy per electron in MH at $D \to \infty$ as a function of $R$ in blue and at $D=3$  in green. } 
    \label{figure_corr_ene_D3__MH}
\end{figure}

In Fig.\ref{figure_corr_ene_D3__MH}, we plot the correlation energy of each electron in metallic hydrogen as a function of $R$ at $D=3$ and compare with the $D=\infty$ result. 

In Fig. \ref{figure_corr_ene_D3__MH}, we see that the correlation energy per electron ($\epsilon_{corr}$) at $D \to \infty$ is bounded by $\epsilon_{corr}$ at $D =3$.  Although at $D=3$, there is no concept of the dihedral angles between electrons, because the electrons are not localized anymore. We shall calculate the corresponding area difference in MH at three dimensions in the next section.

\section{\label{sec:correlation energy in three dimension} Correlation energy and the surface area difference in the three-dimension}

In the previous sections, we established that the correlation energies are bounded by the area differences for N-electron atoms and also for metallic hydrogen at  the large-$D$ limit. In three dimensions, the picture is different, because at $D=3$ the electrons are not localized compared to the  $D \to \infty$ limit. To establish the validity of the area law and correlation energy, we consider the helium atom in three-dimensions. The two $1s$ electrons in the helium atom at $3D$ are in  spherical orbitals with an average electronic radius $\langle  r \rangle $. The average radius changes from  $\langle  r_{HF} \rangle $ (in HF approximation) to $\langle  r_{exact} \rangle $ (with inter-electronic correlation).  The area difference is the difference between the two spherical surfaces with radii $\langle  r_{HF} \rangle $ and $\langle  r_{exact} \rangle $ respectively.

In the Hartree-Fock approximation, the average electronic radius \cite{osti_4553157, boyd1977radial} is given by $\langle  r_{HF} \rangle = 0.92724 $ a.u. On the other hand, with inter-electronic  correlation, a very accurate value of $\langle  r_{exact} \rangle =0.92947$ was computed by Thakkar et al. \cite{PhysRevA.15.1}.

The surface area difference between the two spherical orbitals is calculated as
\begin{equation}
\Delta \text{ area } = 4  \pi \left(r_{HF}^2 - r_{exact}^2 \right) = 0.0520.
\end{equation}

Whereas, the correlation energy of the helium atom \cite{davidson1991ground, veillard1968correlation, JIAO20181} in  atomic unit is given by  $|\epsilon_{corr}| = 0.04204$. Therefore, the correlation energy is bounded by the area difference in three-dimensions also.

\begin{figure}[H]
    \centering
    \begin{tikzpicture}[
 image/.style = {text width=0.8\textwidth, 
                 inner sep=0pt, outer sep=0pt},
node distance = 1mm and 1mm
                        ] 
\node [image] (frame1)
    {\includegraphics[width=\linewidth]{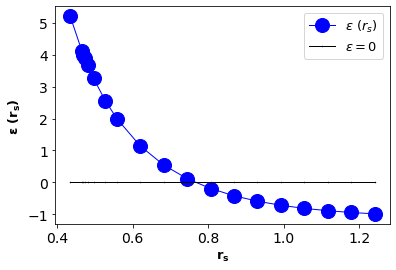}};
\end{tikzpicture}
   \caption{We plot the total ground state energy (in Rydberg unit) per electron in MH at $D=3$ as a function of $r_s$. } 
    \label{figure_total_ene_D3_MH}
\end{figure}

From DFT calculations the ground state energy (in Rydberg unit) per electron in metallic hydrogen in simple cubic lattice at $D=3$  is expressed as \cite{PhysRevB.30.5076}

\begin{equation}
\epsilon (r_s) = \frac{2.21}{r_s^2} - \frac{2.80604}{r_s} -0.13993 - 0.11679 ~\ln~ (r_s), \label{total_energy_MH_at_D_3}
\end{equation}
where $r_s$ is the average atomic radius of each hydrogen atom in MH. 

In Fig. \ref{figure_total_ene_D3_MH}, we  see that the ground state energy becomes positive for $r_s < 0.68$, therefore makes the system is unstable. Therefore, $r_s > 0.68$ can be think of a physically stable region for MH at $D =3$.

Now, if we introduce inter-electronic correlation the average atomic radius in MH will change, i.e. the $r_s$ will change. Therefore, we can think the correlation energy as the change in the ground state energy  due to a change in the atomic radius $r_s$, i.e.

\begin{equation}
\epsilon_{corr} (r_s) = \frac{d \epsilon (r_s)}{d r_s}  \Delta r_s = \epsilon^\prime (r_s) \Delta r_s,
\end{equation}
where $\epsilon_{corr} (r_s)$ is defined in Eq. (\ref{Neece}) and $\epsilon (r_s)$ is defined  in Eq. (\ref{total_energy_MH_at_D_3}). 

On the other hand, the change in the area per atom in MH is given as

\begin{equation}
\Delta \text{(area)} (r_s) = \Delta (4 \pi r_s^2) = 8 ~\pi~ r_s~ \Delta r_s = 8 ~\pi~ r_s~ \epsilon_{corr} (r_s) /  \epsilon^\prime (r_s).  \label{delata_area_for_MH}
\end{equation} 

\begin{figure}[H]
    \centering
    \begin{tikzpicture}[
 image/.style = {text width=0.8\textwidth, 
                 inner sep=0pt, outer sep=0pt},
node distance = 1mm and 1mm
                        ] 
\node [image] (frame1)
    {\includegraphics[width=\linewidth]{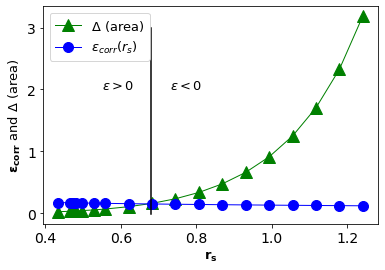}};
\end{tikzpicture}
   \caption{We plot the  $\Delta \text{(area)}$ and $\epsilon_{corr}$ as a function of $r_s$ in blue and green respectively. } 
    \label{figure_corr_ene_andarea_diff_D3_MH}
\end{figure}

In Fig. \ref{figure_corr_ene_andarea_diff_D3_MH}, we plot the $\Delta \text{(area)}$ and $\epsilon_{corr}$ as a function of $r_s$. 

In Fig. \ref{figure_corr_ene_andarea_diff_D3_MH}, we see that the correlation energy per electron in MH at is bounded by the area difference in the physically stable region (i.e. total ground state energy $\epsilon > 0$) in three-dimension also.

\section{\label{sec:Conclusion} conclusion}

Understanding quantum correlation in many-body states of matter, a feature ubiquitous in different problems of physics and chemistry, has gained renewed prominence in recent years. Experimentally, different protocols have been proposed to directly extract spatial correlation functions in phases like in the Mott regime of a Hubbard Hamiltonian \cite{Parsons2016SiteresolvedMO,Mazurenko2017ACF} due to single-site resolution and selective spin removal technique afforded by quantum-gas microscopy in platforms like ultracold-atomic lattices. Such techniques have also been used to study response of magnetic field on an spin-imbalanced 2D lattice of fermions wherein there is a competition to maintain the anti-ferromagnetic checkerboard pattern and alignment with the externally applied field leading to canted lattices \cite{Brown2017SpinimbalanceIA}. Even dynamical correlation from retarded Green's function can be obtained experimentally using Ramsey interferometry as illustrated in \cite{Knap2013ProbingRA} and in quantum simulator of over 100 trapped ions in a Penning trap through carefully designed echo sequences as illustrated in \cite{Garttner2017MeasuringOC}. 

On the other hand, for atomic and molecular systems as studied in traditional electronic structure as well as in this report, different theoretical measures have also been proposed to quantify the extent of deviation from the mean-field Hartree-Fock state like von Neumann entropy with the eigenvalues of the one-particle reduced density matrix (1-RDM)\cite{Pelzer2011StrongCI} or the Cumulant Expansion of the two-particle reduced density matrix (2-RDM)\cite{doi:10.1063/1.478189, Mazziotti2000}. The latter in particular affords many useful characterization including its relationship to eigenvalues of the 2-RDM itself and to long-range order \cite{PhysRevA.92.052502}, its relationship to von Neumann entropy \cite{PhysRevA.97.062502} and its relationship to various orders of cluster amplitudes for a Couple-Cluster (CC) based wavefunction ansatz \cite{kong2011novel}. It has been successfully used to probe electronic correlation in various problems \cite{doi:10.1063/1.2378768, doi:10.1063/1.5140669, doi:10.1063/1.3503766} and very recently in capturing signatures of van der Waals interaction \cite{Werba2019SignatureOV}.

Akin to the latter, in this paper we propose another metric with a simple geometrical insight which can be theoretically used to detect deviation from mean-field behavior and relate the said quantity through an area law to the correlation energy. This is timely as recently various studies \cite{huang2005entanglement, martina2011correlation, dowling2004energy, mohajeri2009information} have shown that the entanglement, a quantum observable, can be used in quantifying  the correlation energy in atomic and molecular systems.  Moreover, a number of studies have shown that the ground state of a local Hamiltonian satisfies an area law and is directly related to the entanglement entropy\cite{ wolf2008area, masanes2009area, de2010ground, michalakis2012stability, anshu2021area, hastings2007area, hastings2007entropy, arad2013area, brandao2013area, landau5143polynomial}. On the other hand, the correlation energy of a system is the difference between the ground state energies in the HF and the accurate calculations. Therefore, the correlation energy is expected  be bounded by the area difference when we go beyond the  Hartree-Fock approximation to an exact representation.  From Fig.\ref{figure_correlation}, we see that the area difference for atomic/ionic systems at large-$D$ limit is a close estimate to the correlation energy of the system. In fact at large-$D$ limit
$\epsilon_{corr} \simeq \Delta(\text{area}) $. In three-dimension,  for helium atom, the area difference  $(0.0520)$ is close to the correlation energy $0.0420$. From Eq.(\ref{delata_area_for_MH}), the average correlation energy of metallic hydrogen in three dimensions can be described as
\begin{equation}
\epsilon_{corr} (r_s) = \alpha ~ \Delta(\text{area}),
\end{equation}
with $\alpha = \epsilon^\prime (r_s) /(8 \pi r_s)$. Combining the above results described in the previous sections we can write an area law for correlation energy as follows
\begin{equation}
\epsilon_{corr} \leq C ~ \Delta(\text{area}),
\end{equation}
with some proportionality constant $C$, which looks similar to Eq.(\ref{area_law_for_entanglement_entropy}), the area law for entanglement entropy.

In summary, we have shown that the correlation energy might be bounded by an area law which is a close resemblance of the  area law conjecture of entanglement entropy. The advantage of this proposal is that we establish a relation between the correlation energy, which is an indirect measure, and the entanglement, which is directly observable, and it is one of the most striking properties of quantum mechanics. Examining the electron correlation in terms of geometry changes between mean field and exact solution might open a new way to observe the correlation effect.

\section*{Acknowledgements}
We would like to thank Dr. Manas Sajjan and Dr. Andrew Hu for  useful discussions and their  comments on the manuscript.  S. K. would like to acknowledge funding by the U.S. Department of Energy (Office of Basic Energy Sciences) under Award No. DE-SC0019215.

\bibliography{ref}

\end{document}